\documentclass[twocolumn,trackchanges,twocolappendix]{aastex631}

\newenvironment{lmss}{\fontfamily{lmss}\selectfont}{\par}
\shorttitle{}
\shortauthors{}
\graphicspath{{./}{fig/}}

\usepackage{amsmath}
\usepackage{graphicx}
\usepackage{siunitx}
\usepackage{diagbox}
\usepackage{multirow}
\usepackage{booktabs}
\usepackage{vcell}
\usepackage{subfigure}
\usepackage{tikz}
\usepackage{colortbl}

\begin{document}

\title{The Polar Vortex Hypothesis: Evolving, Spectrally Distinct Polar Regions Explain Short- and Long-term Light Curve Evolution and Color-Inclination Trends in Brown Dwarfs and Giant Exoplanets
}

\author[0000-0002-6372-8395]{Nguyen Fuda}
\affiliation{Lunar and Planetary Laboratory, University of Arizona, 1629 E. University Boulevard, Tucson, AZ 85721, USA}
\email{fudanguyen@arizona.edu}

\author[0000-0003-3714-5855]{Dániel Apai}
\affil{Steward Observatory, University of Arizona, 933 N. Cherry Avenue, Tucson, AZ 85721, USA}
\affiliation{Lunar and Planetary Laboratory, University of Arizona, 1629 E. University Boulevard, Tucson, AZ 85721, USA}



\reportnum{AAS56053}

\begin{abstract}
Recent studies revealed viewing-angle-dependent color and spectral trends in brown dwarfs, as well as long-term photometric variability ($\sim$100 hr). The origins of these trends are yet unexplained. Here, we propose that these seemingly unrelated sets of observations stem from the same phenomenon: The polar regions of brown dwarfs and directly imaged exoplanets are spectrally different from lower-latitude regions, and that they evolve over longer timescales, possibly driven by polar vortices. We explore this hypothesis via a spatio-temporal atmosphere model capable of simulating time-series, disk-integrated spectra of ultracool atmospheres.
{We study three scenarios with different spectral and temporal components: A null hypothesis without polar vortex, and two scenarios with polar vortices. We find that the scenarios with polar vortex can explain the observed infrared color--inclination trend and the variability amplitude--inclination trend.} The presence of spectrally distinct, time-evolving polar regions in brown dwarfs and giant exoplanet atmospheres raises the possibility that one-dimensional, static atmospheric models may be insufficient for reproducing ultracool atmospheres in detail. 
\end{abstract}


\section{Introduction}\label{sec:intro} 

One-dimensional, static models of brown dwarfs and directly imaged planets have been remarkably successful \citep[e.g][]{saumon_evolution_2008, phillips_new_2020, marley_sonora_2021} and enable major advances in our understanding of the fundamental physical and chemical processes that shape these atmospheres. 
However, recent high-quality data cannot be reconciled with static spherically-symmetric atmospheres: 
Time-resolved spectrophotometry \citep[][]{buenzli_brown_2014} and photometry \citep[][]{metchev_weather_2015} showed that most, if not all, brown dwarfs vary in intensity and spectral features on rotational timescales. In L/T transition brown dwarfs, these changes are attributed to cloud thickness variations \citep[e.g.,][]{radigan_large-amplitude_2012,apai_hst_2013}, which are shaped by planetary-scale waves in equatorial regions shaped by zonal circulation \citep[][]{apai_zones_2017,apai_tess_2021,zhou_roaring_2022,fuda_latitude-dependent_2024, plummer_atmospheric_2024}. These studies show the non-axisymmetric nature of ultracool atmospheres, including temporal evolution.



Other observations suggest an additional process in place. Three lines of observational evidence of brown dwarfs show that the observed properties (spectra and colors) depend on the viewing angle: Firstly, the amplitude of rotational modulations observed in the Spitzer 3.6$\mu m$ band of brown dwarfs was found to be inclination-dependent \citep{metchev_weather_2015, vos_spitzer_2020}. {Secondly, infrared color anomalies of L-T brown dwarfs were also observed to vary with inclination \citep{kirkpatrick_discoveries_2010,metchev_weather_2015,vos_variability_2018, vos_spitzer_2020}}. Thirdly, the silicate index from Spitzer IRS observations of field brown dwarfs \citep{suarez_ultracool_2023-2} indicates that the poles of brown dwarfs might be less cloudy than the equator.
These studies suggest some equator-to-pole variation in the cloud properties of brown dwarfs.

In a seemingly unrelated set of observations, the nearby L/T transition brown dwarf Luhman 16~B \citep{luhman_discovery_2013} displayed a complex lightcurve with distinct variations on rotational timescales (up to 4.7--5.25 hours), and long-term ($\sim$60--125 hours) evolution. \citet{fuda_latitude-dependent_2024} showed that the short-term modulations likely arise from rotational modulation of cloud structures in a zonal circulation-shaped atmospheric (i.e., planetary-scale waves, \citealt[][]{apai_zones_2017}). However, the long-term evolution cannot be explained with rotational modulations and its existence calls for a different mechanism \citep[][]{apai_tess_2021,fuda_latitude-dependent_2024}.

{\citet{zhang_atmospheric_2014} and \citet{showman_atmospheric_2013} predicted two fundamental modes of atmospheric circulation: vortices in heat-transfer-driven regime, and jets/zonal circulation in rotation-dominated regime. Modern GCMs predicted the presence of jets close to the equatorial regions of brown dwarfs, while the high-latitude regions are often dominated by vortices, have a higher degree of vorticity, and redder colors \citep{showman_atmospheric_2020, tan_atmospheric_2021}. Light curve simulations from \citet{tan_atmospheric_2021} (Figure 13) also predicted that the short-period amplitude decreases from equator-to-pole.}

We propose that the inclination-dependent spectrophotometric properties and the long-term photometric evolution of the brown dwarfs are not unrelated but are two manifestations of the same property of ultracool atmospheres.
Specifically, we propose the \textit{polar vortex hypothesis} as follows: While the equatorial- and mid-latitude regions of brown dwarfs and giant exoplanets are shaped by zonal circulation, the polar regions are vortex-dominated and are therefore distinct in appearance and evolve over a longer timescale. In this hypothesis, the combination of jet-dominated bands-and-belts with vortex-dominated polar regions may account for a broad range of observed phenomena and explain the nature of short- and long-term spectrophotometric evolution and inclination-dependent color trends.

{In this paper, we will investigate how the presence of slowly-evolving polar vortices impacts photometry, spectra, and color of ultracool atmospheres.  The paper is structured as follows: In Section \ref{sec:methods}, we describe the atmosphere model and three scenarios of observational tests. In Section \ref{sec:result}, we describe key analysis results from simulated time-series spectra. In Section \ref{sec:discuss} we will discuss our interpretations and the implications of our findings.
}
\section{Methodology: The Atmosphere Model}\label{sec:methods}
\begin{figure*}
\centering
\fbox{
    \parbox[c]{0.8\textwidth}{
    \setlength\lineskip{0pt}

    \includegraphics[width=0.405\textwidth]{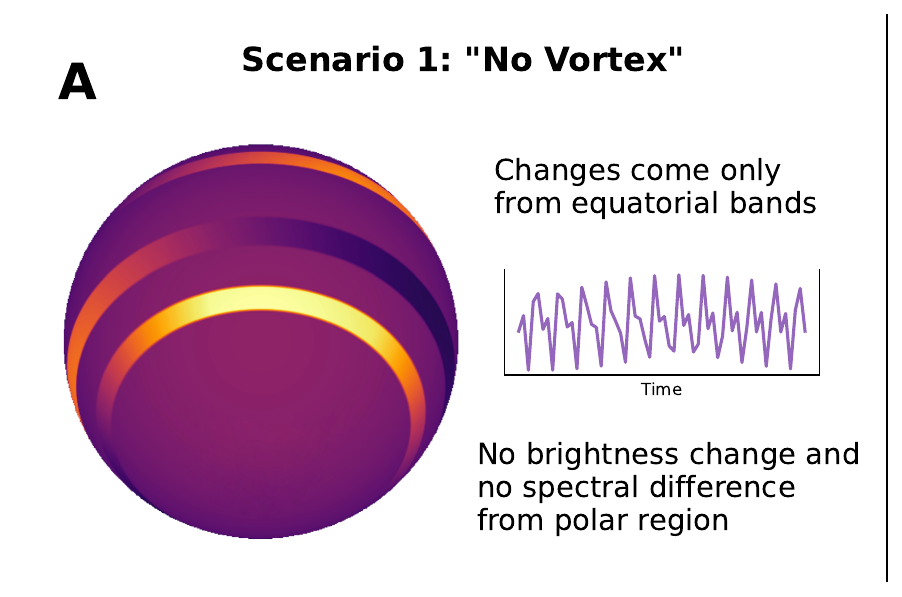}%
    \hspace{0.1cm}
    \begin{minipage}{0.35\textwidth}
        \begin{lmss}
        \fontsize{8pt}{9pt}\selectfont
        \vspace{-4cm}
        \begin{center}
            \textbf{Spatio-temporal Atmosphere Model}
        \end{center}
        \textbf{A:} \textit{`No Vortex'}: null hypothesis, no spectrally distinct poles. (``P" = ``A")
        \newline
        \textbf{B:} \textit{`Evolving Vortex'}: spectrally distinct polar regions, bands change rapidly, poles change slowly.
        \newline
        \textbf{C:} \textit{`Stationary Vortex'}: spectrally distinct polar regions, bands change rapidly \& slowly.
        \newline
        \textbf{D:} Latitudinal spectral model: `A': Ambient, `B': Bands, and `P': Polar.
        \newline
        \textbf{E:} Spectra: `Bands': cloudy, `Polar': less-cloudy, `Ambient': mixture.
        \newline
        \end{lmss}
    \end{minipage}

    \includegraphics[width=0.8\textwidth]{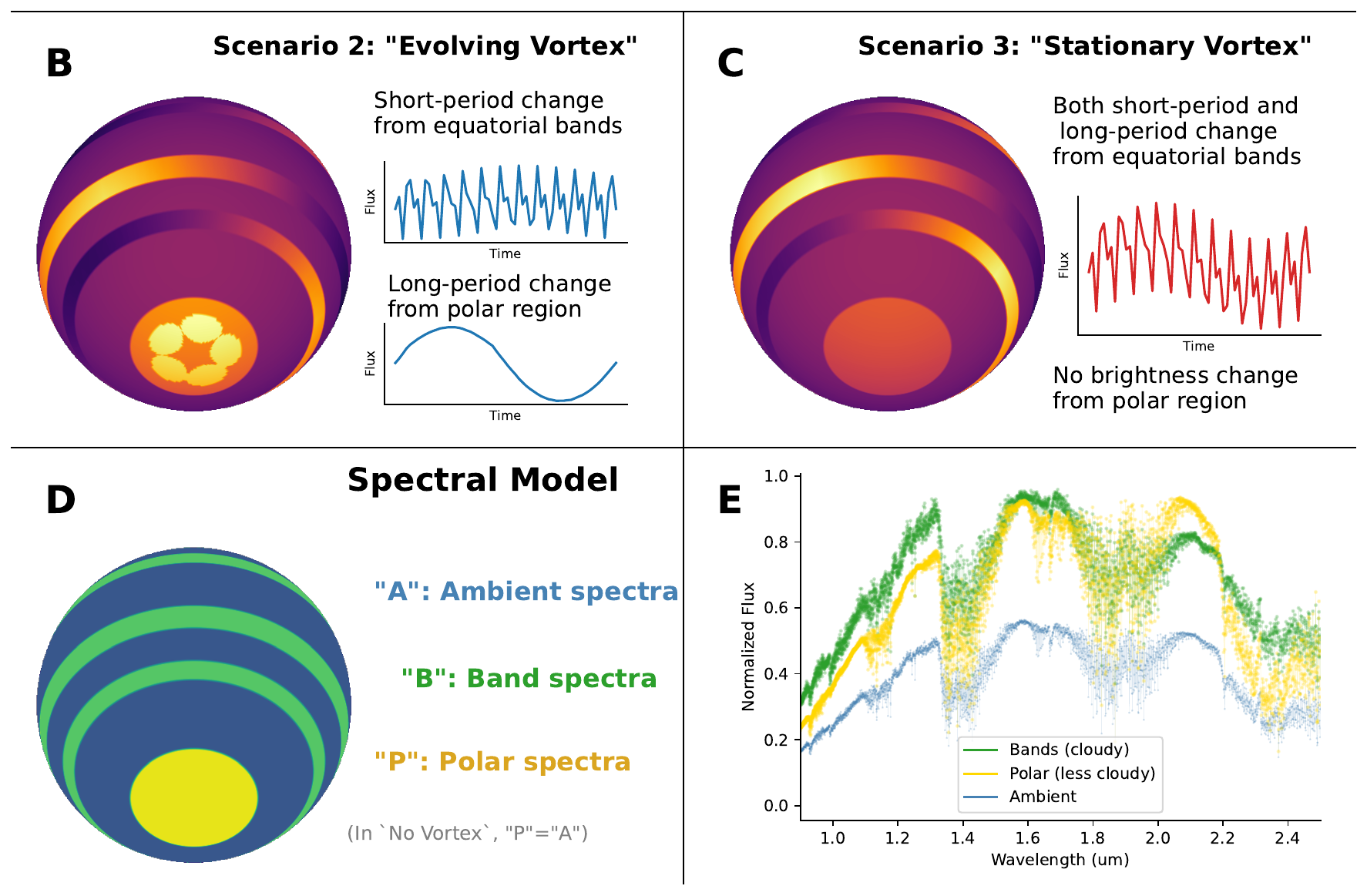}
    
    }}\caption{\textit{Overview of our atmosphere model.} The model is composed of an ultracool atmosphere with spectral types corresponding to two circulation modes: jet-dominated in the equator/mid-latitudes and vortex-dominated in the poles. Sonona-Diamondback cloudy spectra are assigned to the bands, and less-cloudy spectra are assigned to the poles.
    We considered three scenarios: a null hypothesis scenario, `\textit{No Vortex}', where the poles are spectrally indistinguishable from ambient atmosphere; 2) scenario \textit{`Evolving Vortex'} where the polar region is spectrally distinct, and short-period change on rotational timescale comes from the equator/mid-latitudes, and long-period change comes from the pole; scenario `\textit{Stationary Vortex}' where the polar region is spectrally distinct, but all temporal change comes from the equator/mid-latitudes. 
    }
\label{fig:model_overview}
\end{figure*}

We constructed a flexible time-evolving spatio-spectral model. Our model is motivated by previous observations of rotational modulations on brown dwarfs  \citep{apai_tess_2021, fuda_latitude-dependent_2024}, and combines state-of-the-art complex radiative/convective physical-chemical atmospheric models into a simpler multi-component model to study spatio-temporal evolution. Similar models have been used successfully in the past to explain or explain spectrophotometric evolution in brown dwarfs and giant exoplanets \citep{kostov_mapping_2013, apai_hst_2013, karalidi_aeolus_2015, apai_zones_2017}.

Below, we describe the steps of constructing model: 1) Model geometry; 2) Three configurations (``scenarios''); 3) Spectral elements; 4) Simulated observations; and 5) Color and variability amplitude calculation.

\subsection{Model Geometry}\label{subsec:method_geometry}

Our model atmosphere consists of the following surface elements (each with their own assignable) spectra: `Bands', `Polar', and `Ambient' (or `B', `P', `A', see panel A, Figure \ref{fig:model_overview}). 
We represent these in a 2D, 500$\times$500 rectangular brightness array using Python. The array is projected equirectangularly back onto a sphere using \verb|mayavi| \citep[][]{ramachandran_mayavi_2011} to represent a single-pressure spherical atmosphere. The longitudinal extents and relative brightness of the bands are motivated by observations of zonal circulation structures on Jupiter \citep{wong_high-resolution_2020} and Neptune \citep{simon_neptunes_2016}. 

We use latitudes $65^\circ$ North and South as the boundaries for the polar regions, coded `Polar' in the spectral model. We specified 4 different band structures distributed evenly from the equator to mid-latitudes, coded `Bands' and is spatially identical in all scenarios.{ All latitudes outside of `Bands' and `Polar' regions are coded `Ambient' in the spectral model.} Brightness variations in `Bands' are assumed to arise from planetary-scale waves, causing longitudinal cloud thickness difference and when rotated would create flux modulation \citep{apai_zones_2017}. For `Polar', the brightness varies long-term for a whole polar region. {The spectral type `Ambient' does not vary in time in any scenario.}

We prescribe two equatorial-latitude bands as $k=1$ waves, two mid-latitude bands as $k=2$ waves, sinusoidally modulated on periods of 5 and 2.5 hours respectively, as motivated by observation of Luhman 16~B from \cite{fuda_latitude-dependent_2024}. The variability amplitude is $a=5\%$ for all components - this value is chosen to match the Luhman 16 B amplitude. The multi-sinusoidal planetary-scale wave configuration \citep{apai_tess_2021, fuda_latitude-dependent_2024, plummer_atmospheric_2024} is as followed: 
\begin{equation}
    \Sigma^n_i [\: a_i\sin (2\pi / T_i \times t + \phi_i)]
\end{equation}

{Lightcurve shapes and variability amplitudes generated by our model are somewhat sensitive to the specific phases of the waves. To remove potential biases due to specific phase differences, for each inclination angle considered we created 10 iterations with random phases. All the resulting color, variability amplitude, and time-series spectra for a given inclination are averaged over all the phase-randomized iterations (see Sections \ref{subsec:result_incli_evo}, \ref{subsec:discuss_colorVari_inclin}).}

{In summary, the latitudinal extents of the components are: $90^\circ N-65^\circ N \:$ (polar), $45^\circ N-38^\circ N,\: 25^\circ N-15^\circ N,\: 10^\circ S-20^\circ S,\: 33^\circ S-40^\circ S$ (bands), $\: 65^\circ S-90^\circ S$ (polar). All components contain a sinusoidal modulation with periods T = [$60, 2.5, 5, 5, 2.5, 60$] hours with constant variability amplitude $a=5\%$. The phase offsets for each latitude are randomized to remove possible phase-induced patterns. The bands are rotationally modulated, whereas the polar regions uniformly change brightness as a whole.}

\begin{figure*}
    \centering
    \includegraphics[width=0.80\textwidth]{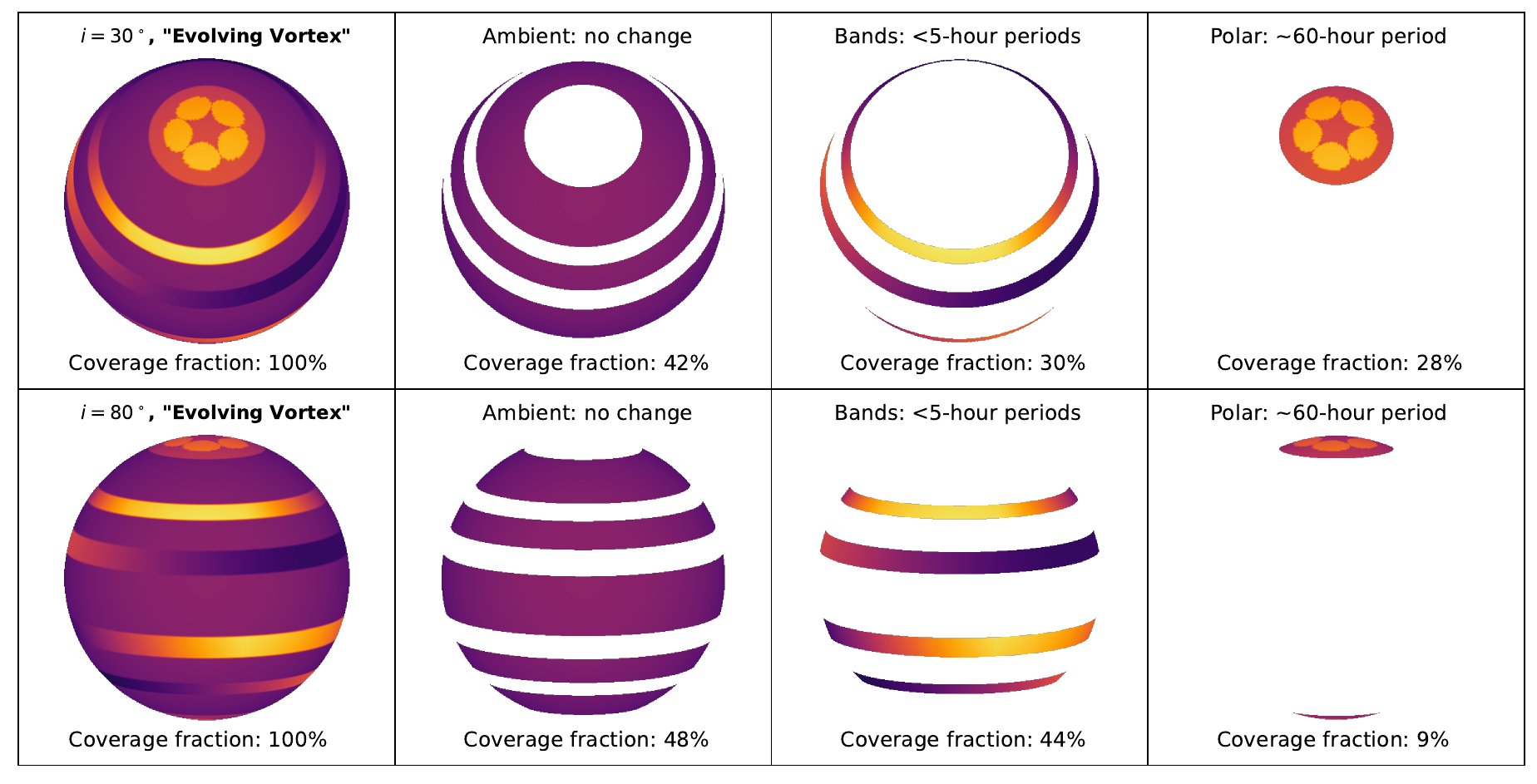}
    \caption{The single-pressure brightness models of the atmosphere, deconstructed of the region type: `Ambient', `Bands', and `Polar'. The coverage fraction of each region with the full disk is shown. The projected area of each region type changes with inclination, which also changes their fractional contribution to the hemisphere-integrated spectra and photometry.
    }
    \label{fig:frac_coverage}
\end{figure*}

\subsection{Three Evolution Scenarios}\label{subsec:method_3scenarios}

In our model, all the scenarios share identical equatorial planetary-scale waves, but their long-period variations and spectral setups differ. The first scenario is \textit{`No Vortex'} (Figure \ref{fig:model_overview}A). This is our null hypothesis for the prediction of polar vortices. In this scenario, the polar regions cannot be spectrally distinguished from the ambient atmosphere. Variability on all timescales comes only from the equatorial bands.

The two other scenarios assume that spectrally distinctive polar vortices exist and the bands, poles, and ambient atmospheres are all spectrally distinct. These scenarios are \textit{`Evolving Vortex'} and \textit{`Stationary Vortex'}.

In the scenario \textit{`Evolving Vortex'} (Figure \ref{fig:model_overview}B), short-period variation comes from rotational modulation of the planetary-scale waves in the bands; long-period variation comes from vortices evolution in the poles. Long-period variability is strongest at pole-on inclination ($i=0^\circ$), and short-period variability is strongest equator-on ($i=90^\circ$).

In the scenario, \textit{`Stationary Vortex'} (Figure \ref{fig:model_overview}C), both short-period and long-period variations come from the bands. 
The polar region is time-invariant in brightness. Here, variability on all timescales will be strongest equator-on ($i=90^\circ$), as the projected area of the bands decreases with more pole-on viewing angle (see Figure \ref{fig:frac_coverage}).

\subsection{Spectra from Atmospheric Models}\label{subsec:method_spectra}

We use different configurations of the Sonora model \citep{marley_sonora_2021, morley_sonora_2024} grids to mimic the different circulation modes with latitude: cloudy for the jet-dominated band region, and less cloudy for the vortex-dominated polar region. The Sonora Diamondback model spectra \citep{morley_sonora_2024} contain different degrees of cloudiness created by the cloud sedimentation efficiency ($f_\text{SED}$) ranging from 1-8 ($f_\text{SED}=1$ most cloudy, $f_\text{SED}=8$ least cloudy). We used $T_\text{eff}=1200 K$, surface gravity $\log g\sim5.0$, $K_{zz}=1$ and metalicity $[M/H]=1$ solar-abundance.

 For the cloudier `Bands' regions, we use $f_\text{sed}=2$ model spectra. For the less-cloudy `Polar' regions, $f_\text{sed}=4$ spectra are used. The `Ambient' regions contain a 1-to-1 mixture of the two spectral types above - {since `Ambient' is unchanging in brightness and is axisymmetric, it contributes no variability to the photometric flux and the spectra and the exact spectral property are less important.} All three spectra types for `'Bands, `Polar', and `Ambient' are plotted in Panel B of Figure \ref{fig:model_overview}) with normalized intensity. The higher-cloudiness `Bands' spectra show a steeper scattering slope at wavelengths shorter than 1.4 $\mu m$, and the less-cloudy `Polar' spectra show deeper molecular absorption features.

 In both scenarios \textit{`Evolving Vortex'} and \textit{`Stationary Vortex'} the polar regions are spectrally different from the bands and ambient atmospheres. However, in scenario \textit{`No Vortex'}, the polar regions and ambient atmosphere are coded with the same spectral type.


\begin{figure*}
    \setlength\lineskip{0pt}
    
    \hspace{1.25cm}\includegraphics[width=0.85\textwidth]{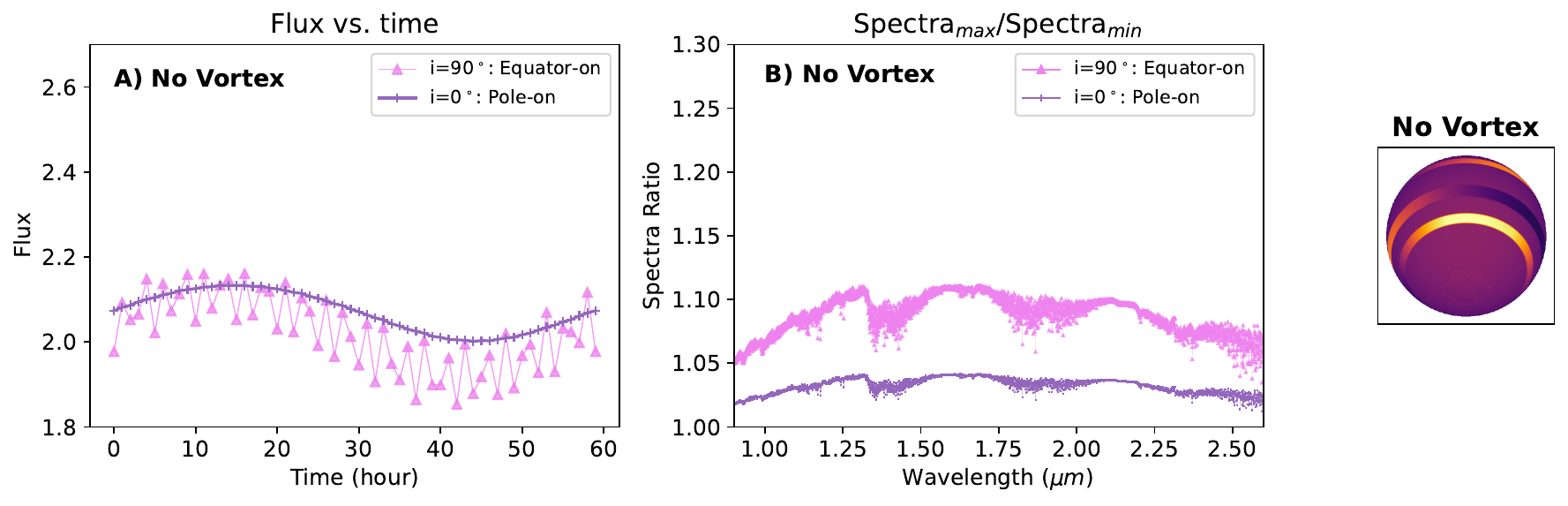}
    
    \hspace{1.25cm}\includegraphics[width=0.8615\textwidth]{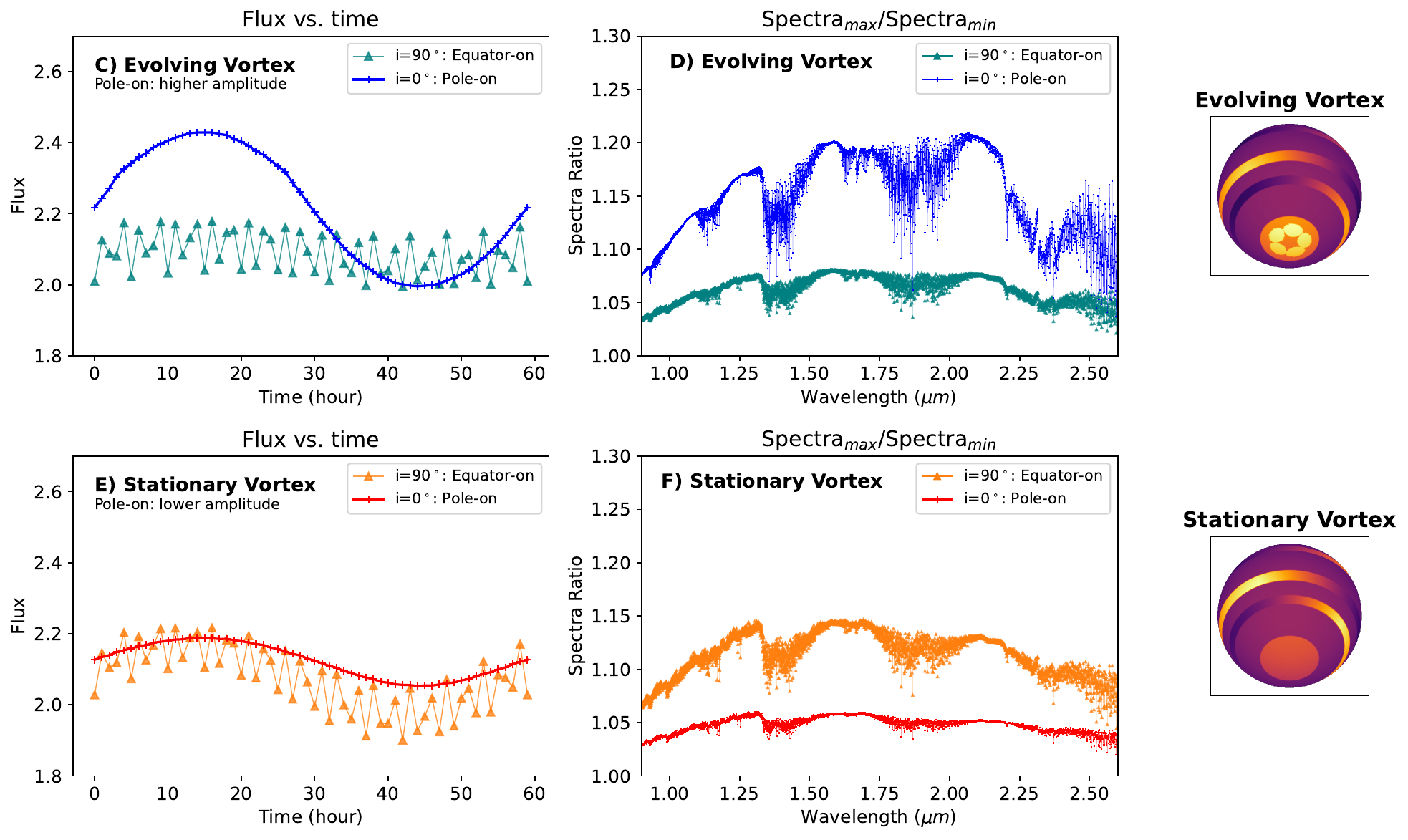}

    \caption{Flux evolution and spectral ratio for scenarios \textit{`No Vortex'} (Panel A, B),  \textit{`Evolving Vortex'} (Panel C, D), and \textit{`Stationary Vortex'} (Panel E, F). The spectral ratio is defined as the ratio of spectra at maximum and minimum flux: $\text{Spectra}_\text{max}/\text{Spectra}_\text{min}$.
    For \textit{`Evolving Vortex'}, the amplitude of flux variation is higher at the pole-on viewing angle.
    In contrast, both \textit{`Stationary Vortex'} and \textit{`No Vortex'} have higher flux amplitude equator-on. 
    This shows that the latitudes where the long-period components are located will affect the observed flux amplitude vs. inclination trend.
    }
    \label{fig:phot_spec}
\end{figure*}

\begin{figure*}
    \centering
    \includegraphics[width=0.80\textwidth]{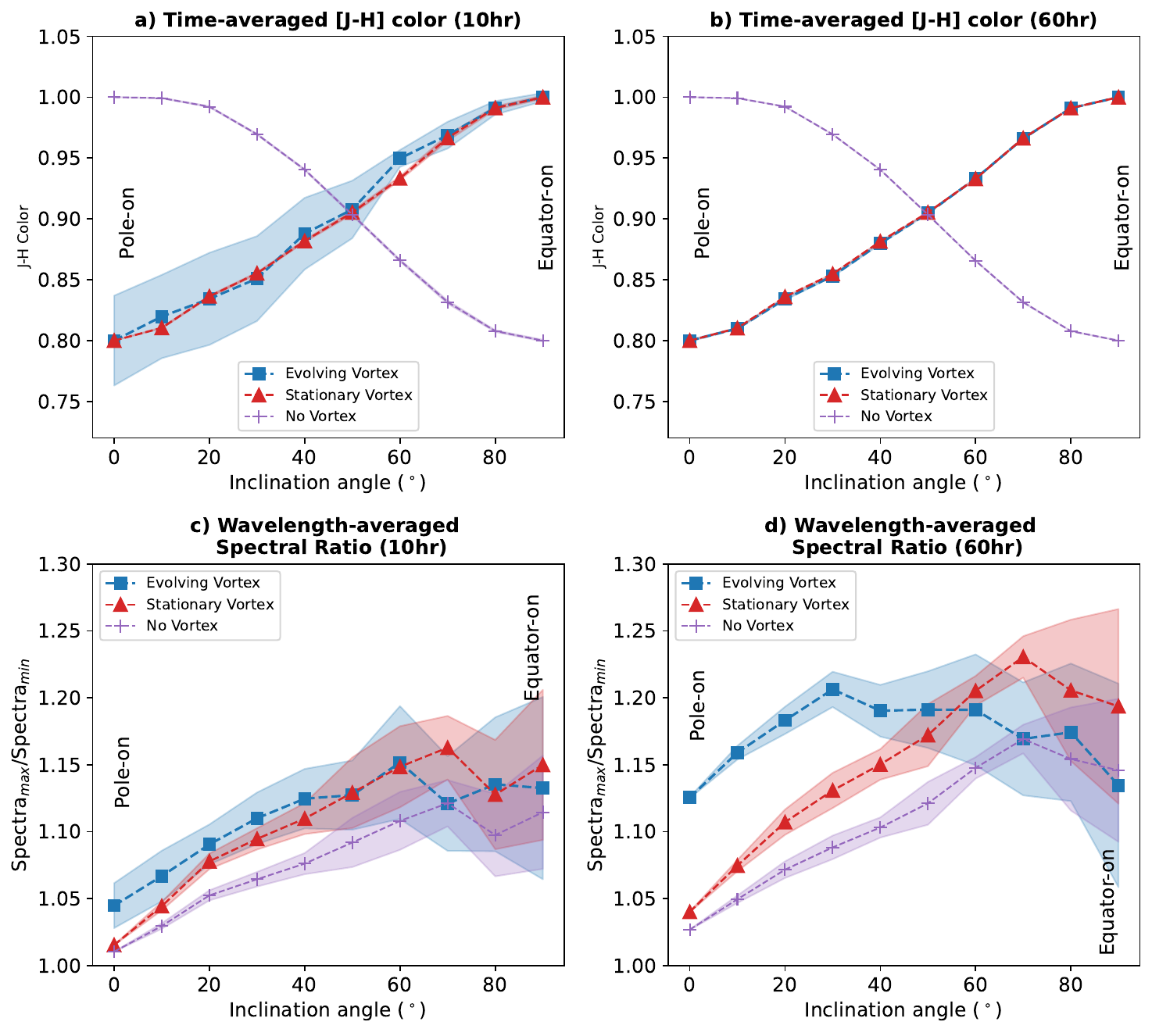}
    
    \caption{{Plots of $J-H$ color (a, b) and wavelength-averaged spectral ratio (c, d) with inclination for each simulated scenario over short-duration (10 hours) and long-duration (60 hours) monitoring. The spectral ratio (defined as $\text{Spectra}_\text{max}/\text{Spectra}_\text{min}$) are averaged over 0.9-2.6 $\mu m$. The shaded regions show $\pm 1 \sigma$ values from 10 phase-randomized iterations.}}
    \label{fig:observables_incli}
\end{figure*}

\subsection{Simulated Spectrophotometric Data Cube}\label{subsec:method_datacube}

After generating the 2D brightness model, we perform an equirectangular projection onto a 3D spherical mesh (equirectangular projection method, \verb|mayavi|\footnote{\href{https://mayavi.readthedocs.io/en/latest/auto/example_spherical_harmonics.html}{\textbf{mayavi}), Spherical Projection.}} with adjustable viewing angle. The 3D atmosphere is then captured as a 2D brightness image for photometry. 

We calculated the disk-integrated spectra as they would emerge from an observation, and simulated an observation campaign with JWST. We calculated disk-integrated spectra for a baseline of 60 hours with a cadence of 1 hour. At each integration, we simulated an $R=2000$ spectra. 

From the photometry timeseries we separately calculated a flux-by-type value. {‘Flux-by-type’ is calculated using the spectral model with three latitudinal encodings: ‘P’ (polar), ‘B’ (bands), and ‘A’ (ambient), generated at the same time as the atmosphere model at each time instance, changing with each inclination. Flux for each component is then calculated by isolating the atmosphere elements with the given encoding.} A spectrum for each type is generated by the relationship: $S(t) = \Sigma_i A_{i,\text{frac}} \times F_i(t) \times S_{i, \text{base}}$, where $A_{i,\text{frac}}$ is the fractional area, $F_i(t)$ the flux and $S_{i,\text{base}}$ the base spectra for each of the three types `Bands', `Polar', `Ambient'. Their contributions to the total flux depends only on the fractional projected area, which in turn depends on the viewing angle (see Figure \ref{fig:frac_coverage}). 

{The final simulated data product is a 3D spectral cube (wavelength--time--inclination angle). Results for each inclination angle contains 10 iterations of the same spectral setup but with different phase randomizations for the time-variable components. Each of those iterations contains one disk-integrated time-series spectra.}

\subsection{Infrared Colors and Variability Amplitudes with inclination}\label{subsec:method_colorVariab}

We calculate the infrared colors from the spectra in the following infrared bands: $J, H, K_s$ centered at $1.10, \: 1.60, \: 2.20 \: \mu m$ respectively, all with $0.1 \: \mu m $ bandpass width. Photometry in each photometric band is calculated by integrating over the wavelength range in the bandpass, multiplied by the transmission curve, and divided by the bandpass width. To calculate the color vs. inclination trend predicted by our models, we calculated the time-averaged [$J-H$] and [$J-K_s$] color throughout the 60-hour observation duration. We define the variability amplitude as half the difference between the maximum and minimum flux over the monitoring duration. {We averaged the values for colors, variability amplitudes, and spectral ratio over 10 phase-randomized iterations at each inclination, and showed the $\pm 1\sigma$ range (results in Section \ref{subsec:result_incli_evo})}.

\section{Results}\label{sec:result}

We will discuss the analysis of the simulated observations from our spatio-temporal model described in Section \ref{sec:methods}. We focus on discussing and comparing the scenarios \textit{`Evolving Vortex'}, \textit{`Stationary Vortex'}, and the null hypothesis scenario \textit{`No Vortex'}. 

\subsection{Temporal and Spectral Evolution}\label{subsec:result_photspec_evo}


{In this section, we focus on only one atmosphere configuration with static phase offsets $\phi$= [$0^\circ, 10^\circ, 150^\circ, -26^\circ, 135^\circ, 0^\circ$] with monitoring timescale of 60 hours. We will describe trends in the flux evolution and spectral ratio for threes scenario using this configuration, and compare pole vs. equator variability and spectra.}

Figure \ref{fig:phot_spec}A shows the photometric flux over time, at pole-on inclination ($i=0^\circ$) and equator-on inclination ($i=90^\circ$) for scenario \textit{`No Vortex'}. Figure \ref{fig:phot_spec}C and \ref{fig:phot_spec}E shows the same for \textit{`Evolving Vortex'} and \textit{`Stationary Vortex'}.  

Figure \ref{fig:phot_spec}B shows the ratio of the spectra at maximum flux to the spectra at minimum flux at pole-on inclination and equator-on inclination for scenario \textit{`No Vortex'}. Figure \ref{fig:phot_spec}D,E shows the same for scenario \textit{`Evolving Vortex'} and \textit{`Stationary Vortex'}. 

For the photometric flux evolution, different locations of time-variable components in each scenario result in different trends. In \textit{`Evolving Vortex'} scenario (Figure \ref{fig:phot_spec}C), the flux amplitude is higher pole-on than equator-on, as the long-period polar component dominates. On the other hand, in the \textit{`Stationary/No Vortex'} scenario, (Figure \ref{fig:phot_spec}A, E), the flux amplitude is higher equator-on than pole-on. Furthermore, for pole-on view, the short-period, variability from the rotationally-modulating bands completely disappears due to geometry.

For the spectral evolution, we observe a similar trend with pole vs. equator. From pole-on to equator-on, the opposite inclination trends occur between scenarios. The wavelength-averaged spectral ratio is larger equator-on than pole-on for \textit{`Stationary Vortex'}/\textit{`No Vortex'} but for \textit{`Evolving Vortex'} it is the reverse.

\subsection{Inclination Dependence: Colors \& Variability Amplitudes}\label{subsec:result_incli_evo}

{In this section, we discuss our results for two different timescales of monitoring: short-duration monitoring (10 hours or 2 rotations) and long-duration monitoring (60 hours or 12 rotations). We will compare and contrast the predicted inclination-dependence of observables (colors, variability amplitudes) between these two timescales.}

{Firstly, the time-averaged infrared colors vs. inclination trends are consistent across the two timescales. The polar vortex scenarios (\textit{‘Evolving Vortex’} \& \textit{‘Stationary Vortex’}) both contain the opposite color-inclination trends compared to the null scenario (\textit{‘No Vortex’})  for both time-averaged [$J-H$] colors. The range of color values is larger over short-duration monitoring because different phase configurations over shorter samples leads to more stochastic results (Figure \ref{fig:observables_incli}a, b).}

{For the \textit{‘Evolving Vortex’} and \textit{‘Stationary Vortex’} scenarios, the overall effect is that larger polar contribution lead to bluer infrared colors. Both scenarios have cloudier Diamondback spectrum with $f_{sed}=2$ in the bands, and less-cloudy $f_{sed}=4$ spectrum in the poles. We found that, the larger the projected areas of the polar regions, the greater the contributions from less-cloudy spectra. The less-cloudy polar spectrum both has a steeper slope around the $J$ and $K_s$ and deeper absorptions at the $H$ band (Figure \ref{fig:model_overview}B).}

{On the other hand, for the \textit{‘No Vortex’} scenario, the constant-brightness polar region is indistinguishable from the ambient atmosphere component, and thus this scenario leads to an opposite color-inclination trend (where the poles are redder than the equatorial region).}

\begin{figure*}
    \centering
    \subfigure[Variability vs. Inclination: Short-duration]
    {\centering
    \includegraphics[width=0.42\textwidth]{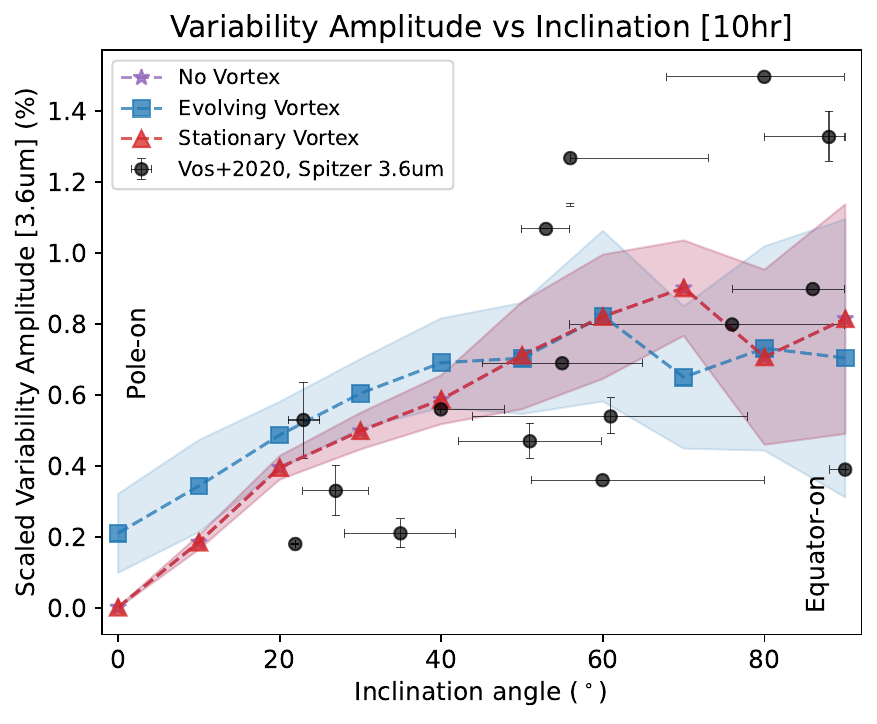}}
    \quad
    \subfigure[Variability vs. Inclination: Long-duration]
    {\centering
    \includegraphics[width=0.42\textwidth]{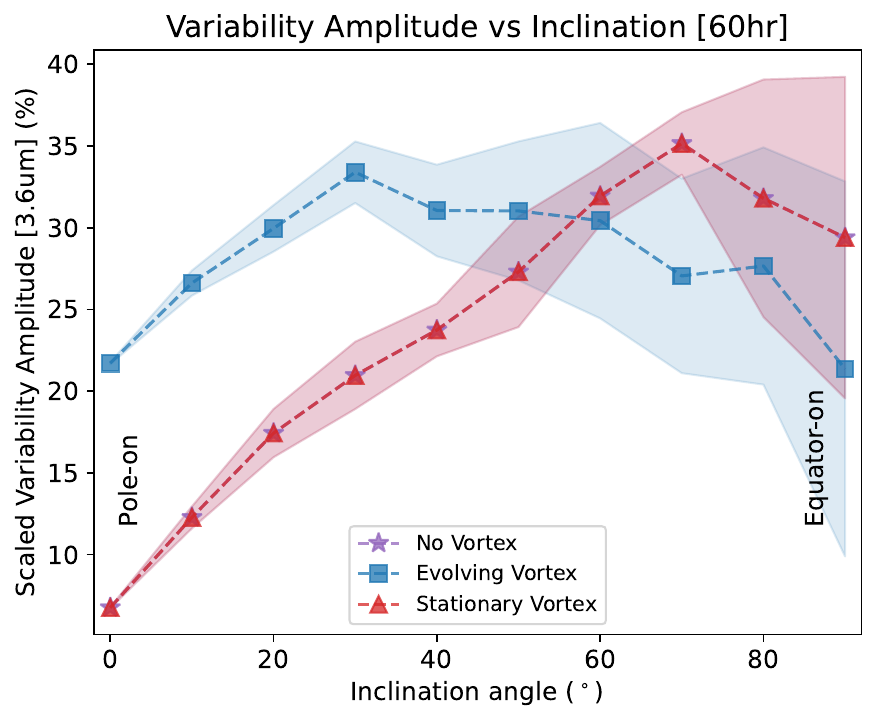}}
    
    \subfigure[Color vs. Inclination: Short-duration]
    {\centering
    \includegraphics[width=0.425\linewidth]{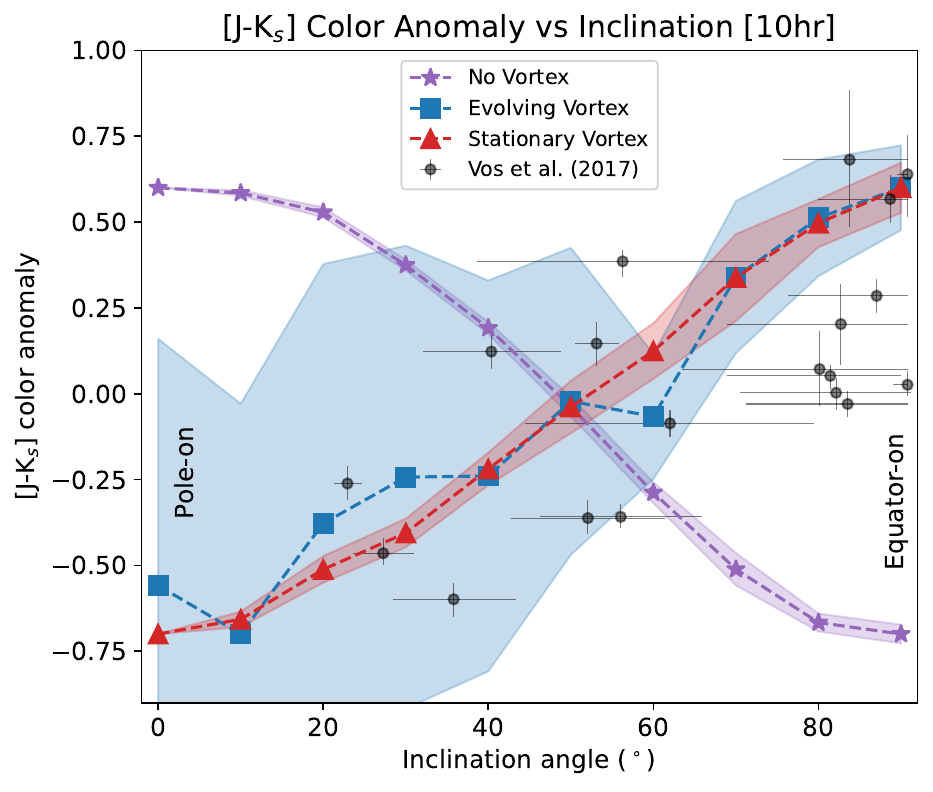}}
    \quad
    \subfigure[Color vs. Inclination: Long-duration]
    {\centering
    \includegraphics[width=0.425\textwidth]{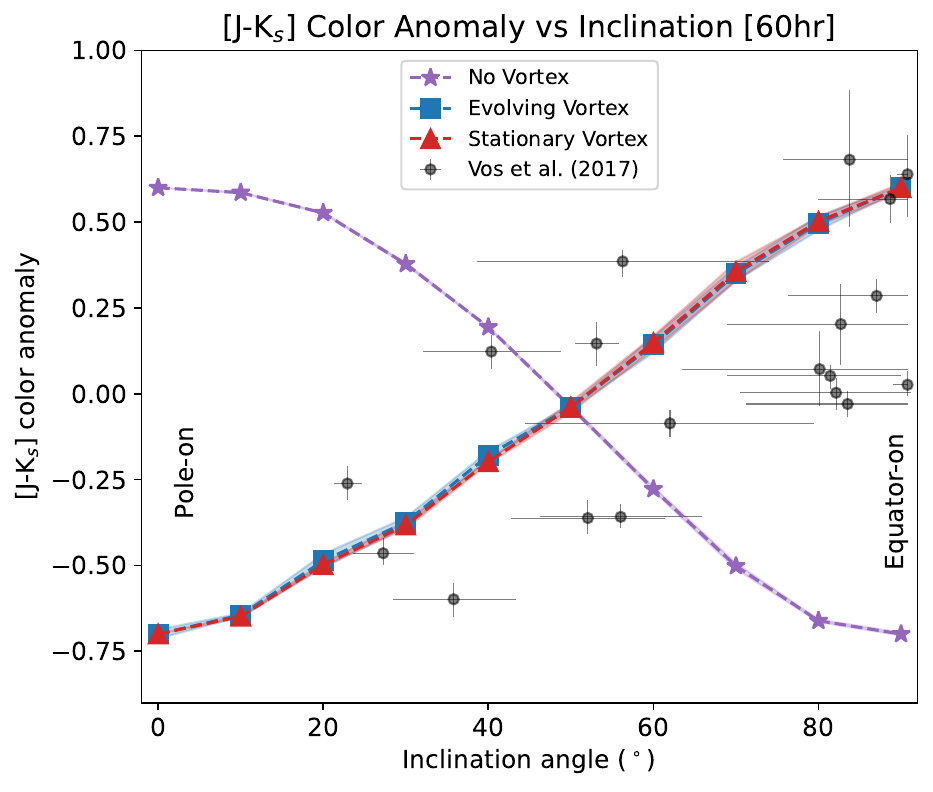}}

    \caption{{ 
    \textit{Panel a,b:} Modeled variability amplitude -- inclination trends for a) Short-duration (10 hours, 2 rotations) and b) Long-duration simulated data (60 hours, 12 rotations). Spitzer 3.6 $\mu$m (few rotations) variability data from \cite{vos_spitzer_2020} overplotted in Panel a. 
    \textit{Panel c, d:} Modeled [\textit{J-K$_s$}] color -- inclination for c) Short-duration and d) Long-duration simulated data. Spitzer color anomaly data from \cite{vos_viewing_2017} overplotted. Both scenarios \textit{`Evolving Vortex'} and \textit{`Stationary Vortex'}  could explain the color-anomaly vs inclination trend. Scenario \textit{`No Vortex'} does not explain the data.}
    }\label{fig:colorVariability_inclination_data}
    
\end{figure*}

{Secondly, the variability amplitude and wavelength-averaged spectra ratio show different inclination-dependence between the two timescales of monitoring (Figure \ref{fig:observables_incli}c, d and Figure \ref{fig:colorVariability_inclination_data}a, b).}

{In the short-duration monitoring, these amplitudes increase from pole to equator viewing angles for all scenarios. Here, the long-term variability amplitude is not captured.} {In the long duration monitoring, the variability amplitudes increase only for the \textit{`Stationary/No Vortex'} scenario, and not for the \textit{`Evolving Vortex'} scenario, which displays inverted U-shape trends with maximum as mid-latitudes.}

{Simulated long-duration monitoring shows an opposite variability-inclination trend between \textit{`Evolving Vortex'} and \textit{`Stationary/No Vortex'} scenarios. The reason is the different origins of long-term variability: Long-term changes are from the polar regions in \textit{`Evolving Vortex'} scenario, and from the low- and mid-latitudes in the \textit{`Stationary/No Vortex'} scenarios. These simulated differences affect the inclination-dependence differently: In Figure \ref{fig:colorVariability_inclination_data}b, the pole-on view of the \textit{`Evolving Vortex'} scenario has significantly higher ($\sim$2x) variability compared to the \textit{`Stationary/No Vortex'} scenarios. Here variability is maximized in the mid-latitudes when both pole and bands are visible.}


\section{Discussion}\label{sec:discuss}

\subsection{Spectral and Photometric Evolution}\label{subsec:discuss_specphot_evo}

{Our results in Section \ref{subsec:result_photspec_evo} demonstrates that the latitude causing long-period evolution significantly impacts inclination trends. Thus, it should be possible to test the presence of slowly varying polar vortices in atmosphere via long-term monitoring.}

The two scenarios \textit{`Evolving Vortex'} and \textit{`Stationary Vortex'} have the variability vs. inclination trend. For \textit{`No Vortex'}, its temporal setup is identical to \textit{`Stationary Vortex'} (Figure \ref{fig:phot_spec}A-C-E). Since the variation amplitude of the `Bands' and `Polar' regions are all equal at 5\%, only the projected areas of each region should determine which component dominates the time series (Figure \ref{fig:frac_coverage}).

For the spectral evolution (Figure \ref{fig:phot_spec}B-D-F), the spectral ratio is larger at the pole-on viewing angle for scenario \textit{`Evolving Vortex'}; but for \textit{`Stationary/No Vortex'} they are larger equator-on. At the same time, the spectral ratio and variation of \textit{`Stationary Vortex'} is larger than that of \textit{`No Vortex'}, since there are no spectral distinctions in the polar regions in the latter scenario. 

Projected area at a given inclination is responsible for the opposite inclination changes. In the pole-on viewing angle, the polar region has the highest projected area and would induce the largest variation coming here while minimizing the area of bands and contribution to the variability. {For the \textit{`Stationary/No Vortex'} scenarios, amplitudes over both the short- and long timescales increase from pole-to-equator. In contrast, compared to the other scenarios, our \textit{`Evolving Vortex'} scenario leads to an opposite variability-inclination trend.}

\subsection{Trends with Inclination of Variability Amplitudes and Infrared Colors}\label{subsec:discuss_colorVari_inclin}

{The simulated observational results vary with the duration of monitoring. Short-duration observation will not capture trends significantly longer than the monitoring baseline. Thus, we compared the inclination-dependence of observables over two different monitoring baselines.}

{First, we compare the 3.6$\mu$m variability amplitude data from \cite{vos_spitzer_2020} with our model. Most objects in the dataset have from one/two to a few rotations measured per visit.} {Figure \ref{fig:colorVariability_inclination_data}a, b shows the modeled variability-inclination trends for a short-duration and long-duration simulated monitoring. We only plotted the Spitzer 3.6$\mu$m data for the short-duration case (Figure 
\ref{fig:colorVariability_inclination_data}a), since the multi-rotations case would not apply to data comparison.}

{The simulated variability-inclination trend from short-duration monitoring agree with observation for all three scenarios. However, for long-duration monitoring, the \textit{`Evolving Vortex'} scenario shows the opposite trend with increasing variability from pole-to-equator. The pole-on variability in scenario \textit{`Evolving Vortex'} is also significantly higher in the long-duration monitoring. Thus, the duration of monitoring has a clear impact on the variability-inclination. This should be a consideration when studying spatial components in time-variable atmospheres.}

Second, we compare the [$J-K_s$] color vs. inclination trends from our model scenarios with field brown dwarf data. \cite{vos_viewing_2017} found a positive correlation between the [$J-K_s$] color and inclination from $i=0^\circ$ to $i=90^\circ$ for $\sim$20 L-T transition brown-dwarfs with measure radial velocities. The [$J-K_s$] color anomaly is the color subtracted by the median color of an object's spectral type. This way, the systematic spectral-type-induced color trend is removed (in the L-T transition, [$J-K_s$] color ranges from very red in the late-L and early-T types, to very blue in the mid-T types).

For both duration of monitorings (Figure \ref{fig:colorVariability_inclination_data}a, b), we plot the [$J-K_s$] color anomaly vs. inclination from \cite{vos_viewing_2017}. We also plot the model [$J-K_s$] color vs. inclination trend with free scaling parameters to match the data for a first-order comparison. For scenarios \textit{`Evolving Vortex'}, \textit{`Stationary Vortex'} the model color vs. inclination trends both match the Spitzer data quite well, where colors get redder from pole-to-equator. For \textit{`No Vortex'}, however, the color trend does not explain the data, in both short-duration and long-duration monitoring. We summarized our results in Table \ref{tab:scenario-summary}.

\subsection{Cloud Thickness Variations across Polar Regions and Bands}\label{subsec:discuss_cloudPolar}

In our model, the higher-cloudiness spectra for `Bands' (Figure \ref{fig:model_overview}E) show a steeper scattering slope at wavelengths shorter than 1.4 $\mu m$, and the lower-cloudiness `Polar' spectra show deeper molecular absorption features. Although these latitude-dependent spectral properties only serve as a first-order approximation of ultracool atmospheres, it has shown color trends consistent with the observation of \citet{vos_viewing_2017} as a function of inclination (Figure \ref{fig:colorVariability_inclination_data}). 

For Solar System giants, cloud decks in the polar regions are thinner and governed by different circulation processes, causing them to have different colors and morphologies than the equator, as observed in the  \citep{west_clouds_2009, zhang_stratospheric_2013, tollefson_neptunes_2019}. For ultracool objects, thickness variation of cloud decks modulated by planetary-scale waves can also create variability on a rotational timescale \citep{apai_zones_2017}; while slowly varying vortices in the polar regions create variability over longer periods. Long-term variations up to hundreds of hours were documented on Luhman 16 AB \citep{apai_tess_2021, fuda_latitude-dependent_2024} --  the only target bright enough for long-term photometric monitoring with TESS. Long-term variations are also believed to be present on multiple other L-T-Y brown dwarfs and directly imaged planets \citep{bedin_hubble_2017}.

\begin{table*}[!htp]
\caption{Summary of the fitting results of $J-K_s$ color anomaly and $3.6\mu m$ variability over the short- and long-duration monitoring for three scenarios. The long-duration simulated monitoring is not applicable to the $3.6 \mu m$ variability amplitude data because the data contains much fewer rotations.}
\label{tab:scenario-summary}
\hspace{-1cm}
\resizebox{\textwidth}{!}{%
\begin{tabular}{|c|l|ll|}
    \hline
    \multirow{2}{*}{\diagbox{Scenario}{Observable}} &
    \multicolumn{1}{c|}{\multirow{2}{*}{\textbf{J-Ks color anomaly}}} &
    \multicolumn{2}{c|}{\textbf{3.6um variablity amplitude}} \\ \cline{3-4} &
    \multicolumn{1}{c|}{} &
    \multicolumn{1}{c|}{\textbf{Short-duration}} &
    \multicolumn{1}{c|}{\textbf{Long-duration}} \\ \hline
    \textit{\textbf{1: `No Vortex'}} &
    \begin{tabular}[c]{@{}l@{}}Bluer equator, redder poles:\\ \textit{does not explain data}\end{tabular} &
    \multicolumn{1}{l|}{\multirow{3}{*}{\begin{tabular}[c]{@{}l@{}}Amplitude increases \\ pole to equator: \\ \textit{explain data of}\\ \cite{vos_spitzer_2020}\end{tabular}}} &
    \begin{tabular}[c]{@{}l@{}}Greater amplitude at \\ equatorial region\end{tabular} \\ \cline{1-2} \cline{4-4} 
    \textit{\textbf{2: `Evolving Vortex'}} &
     \multirow{2}{*}{\begin{tabular}[c]{@{}l@{}}Redder equator, bluer \\ poles: \textit{explain data of} \\ \cite{vos_viewing_2017} \end{tabular}} &
     \multicolumn{1}{l|}{} &
     \begin{tabular}[c]{@{}l@{}}Greater amplitude \\ at polar region\end{tabular} \\ \cline{1-1} \cline{4-4} 
    \textit{\textbf{3: `Stationary Vortex'}} &
    &
    \multicolumn{1}{l|}{} &
    \begin{tabular}[c]{@{}l@{}}Greater amplitude \\ at equatorial region\end{tabular} \\ \hline
\end{tabular}%
}
\end{table*}

\subsection{Polar Vortex}\label{subsec:discuss_polarVor}

{With the \textit{`Evolving Vortex'} scenario, we explored vortex-dominated circulation in the polar region assuming that slowly-evolving vortices drive long-period brightness evolution as suggested by \citet{apai_tess_2021}.} The \textit{`Stationary Vortex'} explored the alternative where short-term rotational modulations could be intertwined with another mechanism changing brightness over a long-term timescale in the bands themselves.

Our results show that the latitudinal origins of a given temporal evolution source will affect the photometric variability amplitude as a function of inclination angle. In particular, long-period components located in the polar regions will have a higher degree of variability amplitude pole-on as represented by the scenario \textit{`Evolving Vortex'}. The reverse is true for the other scenarios, \textit{`Stationary Vortex'} and \textit{`No Vortex'}, where temporal evolution could arise out of jet-dominated equatorial and mid-latitude bands. 

In terms of spectral origins, our result shows that the less-cloudy poles in \textit{`Evolving Vortex'} and \textit{`Stationary Vortex'} can explain the color-anomaly vs. inclination trend from observation of \citet{vos_viewing_2017}. On the other hand, the \textit{`No Vortex'} scenario would not explain the observed color-inclination relationship at all (see Figure \ref{fig:colorVariability_inclination_data}). Our results agree with past studies suggesting that the pole-to-equator color difference relates to cloud thickness: \citet{suarez_ultracool_2023-2} found a positive correlation of the silicate index and inclination, suggesting that the equatorial region of ultracool objects is cloudier than their poles. 

\subsection{Analytical predictions for variability-inclination trends from polar vortex}\label{subsec:discuss_analytical}

\begin{figure}
    \centering
    \includegraphics[width=0.80\linewidth]{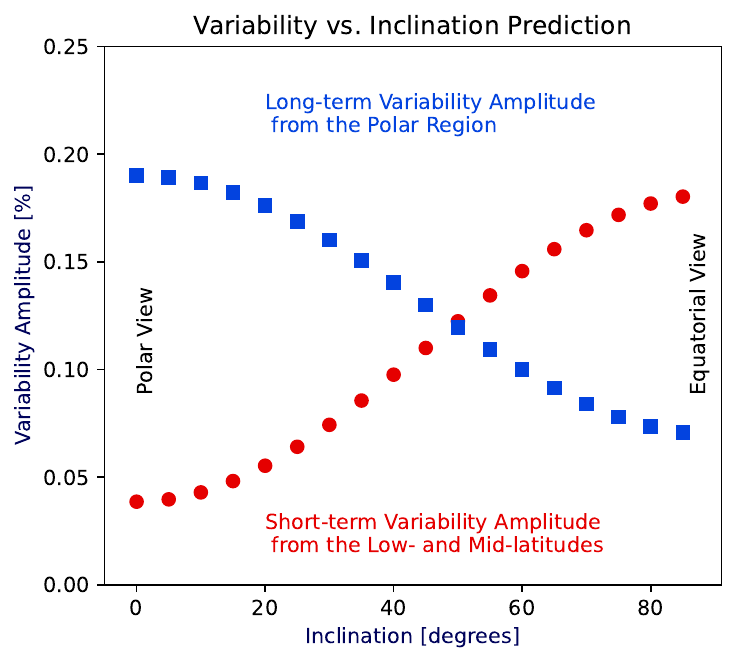}
    \caption{{Analytical variability amplitude vs inclination prediction for atmospheres with long-term variability from the polar region and short-term modulation from the low- and mid-latitudes.}}
    \label{fig:analytical_variabilityIncli}
\end{figure}

{The \textit{`Evolving Vortex'} scenario predicted a different inclination dependence from pole-to-equator as a result of long-term polar variability compared to short-term variability from equatorial bands. We present a simple analytical relationship to explain this finding. The contribution of latitudinal features primarily depends on the visible area of that feature at a given inclination angle. This relationship scales with
$A_i\cos^2[\theta - (90 - \varphi_i)]$ where $\theta$ is a given inclination angle and $A_i$ is the variability amplitude at latitude $\varphi_i$. This means polar features are minimized at equator-on inclination angles and vice versa. In Figure \ref{fig:analytical_variabilityIncli}, we show two atmosphere predictions, one with long-term amplitudes primarily dominant at the poles; and one with short-term amplitudes primarily dominant at the mid-latitudes and the equator. This results in two opposite variability-inclination trends for short- and long-term timescales, consistent with the scenario \textit{`Evolving Vortex'} and comparison with the observed variability-inclination trends and color-inclinationt trends discussed in Section \ref{subsec:discuss_colorVari_inclin}.}

\subsection{Comparison to Solar System Giant Planets}\label{subsec:discuss_solarCompare}

Ultracool atmospheres (i.e. weakly-irradiated brown dwarfs, directly imaged planets) share significant parameters space with the cold Solar system gas giants \citep{showman_atmospheric_2020}. 
Studies of Jupiter, Saturn, Neptune, and Uranus atmospheres have revealed that their polar atmospheres are in stark contrast to the equators. 
Examples of vortex-dominated poles on Solar System giants are abundant: JunoCam images of Jupiter's poles showing the blue-colored multi-cyclone system in its polar tropospheres \citep{orton_first_2017}; Cassini images of Saturn's striking hexagonal polar cyclones \citep{porco_cassini_2003}; ALMA observation showing Neptune's south polar hot spot \citep{tollefson_neptunes_2019}; VLA observation of Uranus's dominant polar cyclone \citep{akins_evidence_2023}. Differences in circulation regimes leading to different appearances could explain the pole-to-equator color evolution on these planets, and potentially the observation of brown dwarfs' color evolution \citep{vos_viewing_2017, suarez_ultracool_2023-2} as well. 

Temporal variation in giant's atmospheres is also well documented. For Jupiter, \citet{ge_rotational_2019} suggested presence of rotational modulations over many wavelengths; \citet{hori_jupiters_2023} suggested torsional oscillation responsible for long-term evolution. For Neptune, rapid changes in brightness were observed on Neptune \citep{simon_neptunes_2016} with latitudinal planetary-scale waves \citep{tollefson_neptunes_2019} modulating the brightness; slower, long-term variation over a 20-year timespan \citep{chavez_evolution_2023}. Past brown dwarf light curves have demonstrated similarly rapid variation, and also slow variation over dozens of rotations, namely Luhman 16AB \citep{apai_tess_2021, fuda_latitude-dependent_2024}. 

Observation of Jupiter suggested that vortice morphologies vary over time \citep{adriani_clusters_2018, wong_high-resolution_2020, mitchell_polar_2021}. Vortices emerging and disappearing from the polar regions \citep{li_modeling_2020} could create long-term changes in the atmosphere. Periodic, multi-year changes have been documented \citep{orton_unexpected_2023} via infrared observations of Jupiter, pointing to variation of tropospheric temperature in the latitudinal bands. 

In summary, clear evidence of pole-to-equator spectrotemporal differences is present on Solar System giant planets. Our model scenarios are designed to be consistent with this fact. Since these pole-to-equator differences might also exist on ultracool atmospheres, the Solar System "ground-truth" will help interpret observations on brown dwarfs and directly imaged planets. The model scenarios we presented above can help generate observational tests to search for polar vortex on ultracool atmospheres. 

\section{Conclusions}\label{sec:conclusion}

Recently, long-term photometric variations were observed in brown dwarfs. Separate studies have also shown that a near-infrared color--anomaly and inclination trend are present in brown dwarfs. Here, we propose a hypothesis in which the two seemingly unrelated observations are different manifestations of the same phenomenon: 
The polar regions of brown dwarfs and weakly irradiated giant exoplanets are vortex-dominated and therefore have distinct spectral properties and temporal evolution. Our hypothesis is supported by observational evidence that the polar regions of both brown dwarfs and Solar System giants (Jupiter and Saturn) have different colors and are less cloudy than their equators. In addition, some brown dwarf and giant exoplanet GCM simulations show vorticity and long-term photometric variations in the polar regions. To explore the testability of the polar vortex hypothesis, we built an evolving spatial-temporal model for an atmosphere. Our model can include time-evolving bands (which could be the results of zonal circulation) and time-evolving polar regions. 

The key findings of our study are as follows:

\begin{enumerate}
    \itemsep0em 
    \item {We considered three scenarios (\textit{`No Vortex'}, \textit{`Stationary Vortex'}, and \textit{`Evolving Vortex'}). We evaluated if the modeled observations are consistent with the observed color anomaly vs. inclination trends and timescale modulations. }
    \item {We found that the null hypothesis \textit{`No Vortex'} does not explain the observed [$J-K_s$] color anomaly vs. inclination trend. }
    \item {In contrast, we found that  the (\textit{`Evolving Vortex'} and \textit{`Stationary Vortex'}) scenarios can explain the observed [$J-K_s$] color anomaly vs. inclination trends, becoming redder from pole-to-equator.}
    \item {For simulated short-duration monitoring, we found that all scenarios explain the variability-inclination trends increasing from pole-to-equator seen in Spitzer 3.6$\mu$m, as short-period bands are most prominent around the equatorial region in our model.}
    \item {For simulated long-duration (12 rotations) observations, we found that scenario \textit{`Evolving Vortex'} (where the polar regions are slowly evolving) shows the opposite variability-inclination trends compared to either \textit{`Stationary Vortex'}/\textit{`No Vortex'} (where the bands are slowly evolving). } \end{enumerate}

We suggest that brown dwarfs and giant exoplanets have multi-component atmospheres with viewing-angle-dependent spectra and spectrophotometric evolution. {In summary, our study shows that a less-cloudy polar regions can explain the observed color-inclination trends. We also predict that the variability-inclination trend in long-duration monitoring will be the opposite between the \textit{`Evolving Vortex'} and \textit{`Stationary/No Vortex'} scenario.} Our polar vortex hypothesis is testable via multi-visit, long-term space-based monitoring with telescopes such as JWST, Ariel, or Pandora.

\section*{Acknowledgements}
Support for the programs HST-GO-16296 and JWST-AR-4927 were provided by NASA through grants from the Space Telescope Science Institute operated by the Association of Universities for Research in Astronomy, Incorporated. We acknowledge support from the Pandora SmallSat mission operated by the Goddard Space Flight Center through grants provided by the NASA Pioneers Program.

\section*{Source Code}
\dataset[The source code is publicly available via Zenodo.]{https://zenodo.org/records/13852273}

\bibliography{main.bib}{}
\bibliographystyle{aasjournal}
\end{document}